\documentstyle[preprint,aps]{revtex}

\begin{document}

\title{FORM INVARIANCE OF DIFFERENTIAL EQUATIONS IN GENERAL RELATIVITY}

\author{Luis P. Chimento\footnote {Fellow of the
Consejo Nacional de Investigaciones Cient\'{\i}ficas y T\'ecnicas.}\\
{\it Departamento de F\'{\i}sica,
Facultad de Ciencias Exactas y Naturales,\\
Universidad de Buenos Aires,
Ciudad Universitaria, Pabell\'{o}n  I,
1428 Buenos Aires, Argentina.
E-mail: chimento@df.uba.ar}}

\maketitle

\begin{abstract}

\noindent Einstein equations for several matter sources in Robertson-Walker
and Bianchi I type metrics, are shown to reduce to a kind of second order
nonlinear ordinary differential equation $\ddot{y}+\alpha f(y)\dot{y}+\beta
f(y)\int{f(y)\,dy}+\gamma f(y)=0$. Also, it appears in the generalized
statistical mechanics for the most interesting value $q=-1$. The invariant
form of this equation is imposed and the corresponding nonlocal transformation
is obtained. The linearization of that equation for any $\alpha$, $\beta$ and
$\gamma$ is presented and for the important case $f=by^n+k$ with $\beta=\alpha
^2\frac{n+1}{(n+2)^2}$ its explicit general solution is found. Moreover, the
form invariance is applied to yield exact solutions of same other differential
equations.
 
\end{abstract}

\noindent Short title: FORM INVARIANCE OF DIFFERENTIAL EQUATIONS


\newpage
\section{Introduction}

Exact solutions of the Einstein equations are difficult to obtain due to their
nonlinear nature. There exist several interesting physical problems where the
Einstein field equations for homogeneous, isotropic and spatially flat
cosmological models with no cosmological constant \cite{visco}-\cite{viale}
and for a time decaying cosmological constant \cite{Reut}, or Bianchi I-type
metric \cite{juan} with a variety of matter sources, reduce to particular
cases of the second order nonlinear ordinary differential equation

\begin{equation}
\label{1}
\ddot{y}+\alpha f(y)\dot{y}+\beta f(y)\int{f(y)\,dy}+ \gamma f(y)=0,
\end{equation}

\noindent where $y=y(x)$, $f(y)$ is a real function and the dot means
differentiation with respect to $x$. $\alpha$, $\beta$ and $\gamma$ are
constant parameters.

Recently, it was shown that some galactic models of astrophysical relevance,
when investigated with the ``generalized'' Statistical Mechanics \cite{t1},
can be exactly described by solutions to the Boltzmann equations that maximize
the Generalized Tsallis Entropy for $q=-1$ \cite{t8}, and it was found that
the corresponding probability distribution function satisfies (\ref{1})
\cite{flavia}.

It is believed that quantum effects played a fundamental role in the early
Universe. For instance, vacuum polarization and particle production arise from
a quantum description of matter. It is known that both of them can be modeled
in terms of a classical bulk viscosity \cite{Hu}. Using the relativistic
second-order theory of non-equilibrium thermodynamics called Extended
Irreversible Thermodynamics developed in \cite{Pav82} \cite{Jou93}, it was
considered a homogeneous isotropic spatially-flat universe, filled with a
causal viscous fluid whose equilibrium pressure obeys a $\gamma$-law equation
of state, while the transport equation of the viscous pressure is

\begin{equation}
\label{1.6}
\sigma+\tau\dot\sigma=-3\zeta H-\frac{1}{2}\epsilon\tau\sigma
\left(3H+\frac{\dot\tau}{\tau}-\frac{\dot\zeta}{\zeta}-
\frac{\dot T}{T}\right).
\end{equation}

\noindent with $\epsilon =0$ \cite{Pav}. Following \cite {Bel79} for $m=1/2$,
it was shown in \cite{visco} that the expansion rate satisfies a modified
Painlev\'e-Ince equation that has the form of (\ref{1}) with $f(y)=y$ and
$\gamma =0$.


Cosmological models with a viscous fluid source have been studied using the
full causal irreversible thermodynamics with the full version of the transport
equation for the bulk viscous pressure \cite{vicenc} \cite{visco3}
\cite{viale}. Relating the equilibrium temperature $T$ with the energy density
in the simplest way to guarantee a positive heat capacity, it was shown that
the expansion rate satisfies (\ref{1}) for $m=1/2$, with $f(y)=y^{-1/r}$ and
$\gamma =0$ \cite{visco3}. Also, the early time evolution of a dissipative
universe, leads to an equation for the expansion rate that has the form
(\ref{1}) \cite{visco2},\cite{Zak}, in the relaxation dominated regime.


Another interesting example appears when an anisotropic universe, described by
a Bianchi type I metric, is driven by a minimally coupled scalar field with an
exponential potential. The Klein-Gordon equation for the scalar field and the
Einstein equations for the metric are expressed in term of the semiconformal
factor $G$ and their derivatives \cite{AFI}. Then, the solutions of this
equation set can be obtained if one is able to solve the following Einstein
equation for $G$,

\begin{equation}
G\frac{\ddot G}{\dot G}+(c-1)\dot G+ \frac{c_1}{\dot G}=c_2,
\label{first}
\end{equation}

\noindent which, making the substitution $G=y^{1/c}$ (\ref{first}) becomes
(\ref{1}) \cite{juan}. A similar result is obtained in the particular case
when the Bianchi type I metric reduces to a flat Robertson-Walker space-time
\cite{Luis}.


From the Generalized Tsallis Entropy, defined as \cite{t1}

\begin{equation}
\label{f1}
S_q=k(q-1)^{-1}\sum\limits_i(p_i-p_i^q),
\end{equation}

\noindent it can be constructed the generalized Statistical Mechanics where
$k$ is a positive constant, $q$ is a real number that characterizes the
statistic and the sum is made over all the microscopic configurations whose
probabilities are $p_i$. It leads to the conventional Boltzmann-Shannon
statistic in the limit $q\rightarrow 1$ and it is found to be a good framework
to study astrophysical problems, as the Generalized Freeman Disk \cite{t11}
and Kalnajs oscillations of a slab of stars \cite{t12}. Taking the generalized
Fisher information for Tsallis Statistics \cite{inf}

\begin{equation}
\label{f2}
I_q=\langle \left(\frac{\frac d{dx}f_d}{f_d(x)}\right) ^2\rangle,
\end{equation}
 
\noindent where $f_d(x)$ is the probability distribution function, and solving
the variational problem in order to find the distribution function that
maximizes the Fisher information, a differential equation of type (\ref{1}) is
obtained for $y=\dot{f_d}/f_d$, where $f(y)=y$, $\alpha=(2q-1)$,
$\beta=\frac 12 q(q-1)$ and $\gamma=0$ \cite{flavia}. For relevant physical applications the
most interesting value of the statistic parameter is $q=-1$ \cite{t8}, in this
case the above equations can be solved explicit and the general solution will
be given in section 3.

Thus, it turns out to be of great interest to analyze (\ref{1}) from the
physical and mathematical point of view. The paper is organized as follows, in
section II we introduce an invariant form and use it to reduce (\ref{1}) to a
linear, inhomogeneous ordinary second order differential equation with
constant coefficients, by means of a nonlocal transformation. Then, its
parametric general solution is given. In section III we extend the nonlocal
transformation and find the explicit general solution of a modified
Painlev\'e-Ince equation for $\beta=1/9$ \cite{Ince}. In section IV we use the
nonlocal invariance to obtain a new class of differential equations for which
the general solution is found. In section V the conclusions are stated.

\section{Form invariance}

The differential equation (\ref{1}), which appears in several interesting
physical problems, has been solved and studied in particular cases using
nonlocal transformations as it was previously stated. To investigate (\ref{1})
we write it in invariant form

\begin{equation}
\label{6}
\frac{\ddot{y}}{f(y)}+\alpha \dot{y}+\beta \int{f(y)\,dy}+ \gamma =
\frac{{\overline y}\,''}{\overline f(\overline y)}+
\overline \alpha \,\overline y{\,}'+
\overline \beta \int{\overline f(\overline y)\,d{\,}\overline y}+
\overline \gamma,
\end{equation}

\noindent under the nonlocal transformation group defined by the
transformation

\begin{equation}
\label{7}
\beta f(y)\,dy=\overline\beta\,\overline f(\overline y)\,d{\,}\overline y,
\end{equation}

\begin{equation}
\label{8}
\frac{\beta}{\alpha}f(y)\,dx=
\frac{\overline \beta}{\overline \alpha}\overline f(\overline y)\,d{\,}
\overline x,
\end{equation}

\begin{equation}
\label{9}
\frac{\beta}{\alpha^{2}}=\frac{\overline \beta}{\overline \alpha\,^{2}},
\end{equation}

\begin{equation}
\label{9.1}
\beta c+\gamma=\overline\beta\overline c+\overline\gamma,
\end{equation}

\noindent where $\overline f(\overline y)$ is a real function of $\overline
y=\overline y(\overline x)$, the prime indicates differentiation with respect
to $\overline x$. $\overline\alpha, \overline\beta, \overline\gamma$ are
constant parameters and $c(\overline c)$ is an integration constant provided
by the integral on the left(right) hand side of (\ref{6}). By invariant
form we mean that the left hand side of (\ref{6}) transforms into the right
hand side under the nonlocal transformation defined by
(\ref{7}-\ref{9.1}) for any functions $f,\overline f$. The parameters
$\alpha$, $\beta$, $\gamma,$ $\overline\alpha$ and $\overline\beta$ satisfy
(\ref{9}-\ref{9.1}).

The form invariance group can be used to linearize (\ref{1}). In fact,
taking the function $\overline f(\overline y) =1$, $\overline \alpha =\alpha$,
$\overline \beta =\beta$ and $\overline\gamma=\gamma$ (this means $\overline
c=c$) in the invariant form (\ref{6}) and the transformation
(\ref{7}-\ref{9.1}), they become

\begin{equation}
\label{10.1}
\frac{\ddot{y}}{f(y)}+\alpha \dot{y}+\beta \int{f(y)\,dy}+\gamma=
{\overline y}{\,}''+\alpha\overline y{\,}'+\beta{\,}\overline y+\beta c+
\gamma,
\end{equation}

\begin{equation}
\label{10}
\overline y=\int{f(y)\,dy}, \qquad{\overline x=\int{f(y)\,dx}}.
\end{equation}

\noindent Without loss of generality we choose $\overline c=c=0$. So, if the
invariant (\ref{10.1}) vanishes, then, (\ref{1}) transforms into

\begin{equation}
\label{11}
\overline y{\,}''+\alpha \overline y {\,}'+\beta{\,}\overline y+\gamma =0,
\end{equation}

\noindent under the transformation of variables (\ref{10}). This is a linear,
second order ordinary differential equation with constant coefficients . Its
general solution is

\vskip .5cm
\noindent a) $\beta\ne \frac{\alpha ^2}{4}$

\begin{equation}
\label{12}
\overline y= c_1\exp{\left(\lambda_{1}\overline x\right)}
+c_2\exp{\left(\lambda_{2}\overline x\right)}-\frac{\gamma}{\beta},
\end{equation}

\noindent where $\lambda_1$ and $\lambda_2$ are the roots of the
characteristic polynomial of (\ref{11}). We indicate the integration constants
with $c$, $c_1$,...$c_n$ and $\overline c$, $\overline c_1$,...$\overline
c_n$.

\vskip .5cm
\noindent b) $\beta =\frac{\alpha ^2}{4}$

\begin{equation}
\label{14}
\overline y = (c_1+c_2\overline x)\exp {\left(-\frac{\overline x}{2}\right)}-
\frac{\gamma}{\beta}.
\end{equation}

\noindent The real solutions can be classified as follows (we also assume
that $\alpha$, $\beta$ and $\gamma$ are real). For $\alpha >0$ and
$\beta < \frac{\alpha ^2}{4}$ we have two real, negative roots for a strong
damped solution. For $\beta=\frac{\alpha ^2}{4}$ we have a double-negative
root for a critically damped solution. For $\alpha >0$ and $\beta >\frac{\alpha
^2}{4}$ we have two complex roots with negative real parts for a weakly damped
solution. For the case $\alpha<0$ growing solutions occur.

The transformation of variables (\ref{10}), relates the general solution of
(\ref{1}) with $\overline y (\overline x)$ through (\ref{12}). We find that

\begin{equation}
\label{15}
y=y(\overline y(\overline x)),
\end{equation}

\begin{equation}
\label{16}
x=\int{\frac{1}{f(y(\overline y(\overline x)))}\,d{\,}\overline x}.
\end{equation}

\noindent are the parametric equations for $x$ and $y$
in terms of $\overline x$. In the particular case $f(y)=y$ we have shown that
a class of nonlinear modified Painlev\'e-Ince equation can be transformed
into a linear second order ordinary differential equation by a nonlocal
transformation.

The theory introduced by Lie considers the invariance of the differential
equations under point transformations. He showed that the one-dimensional free
particle equation has the eight-dimensional SL(3,R) group of point
transformations. This is the maximum number of symmetry generators for a
second-order differential equation of the form \cite{stephani}

\begin{equation}
\label{16.1}
\ddot y+h(\dot y,y,x)=0.
\end{equation}

\noindent In our case (\ref{1}) has the form of (\ref{16.1}). Then, it has
eight or less point symmetries. However, it becomes (\ref{11}) under the
transformation of variables (\ref{10}) and can be cast into the free particle
equation by a local point transformation. So, (\ref{11}) has always eight
symmetry generators. We conclude this section observing that the nonlocal
transformation (\ref{7}-\ref{9.1}) changes the number of symmetry generators
for the class of differential equations (\ref{1}) and the physics contained in
the original problem.

\subsection{The nonconstant parameters case}

Here we allow the parameters in (\ref{1}) and in the transformation
(\ref{7}-\ref{9.1}) to be functions of the independent variable, that is,
$\alpha =\alpha (x)$, $\beta =\beta (x)$ and $\gamma =\gamma (x)$. In order to
preserve the form (\ref{1}) we choose $\overline\alpha(\overline
x)=\alpha(\overline x)$ and $\overline\beta(\overline x)=\beta(\overline x)$.
In this case, the invariant form (\ref{6}) reads

\begin{equation}
\label{2b.1}
\frac{\ddot{y}}{f(y)}+\alpha(x)\dot{y}+\beta(x)\int{f(y)\,dy}+\gamma(x)=
\frac{{\overline y}\,''}{\overline f(\overline y)}+\alpha(\overline x)\overline y{\,}'+
\beta(\overline x)\int{\overline f(\overline y)\,d{\,}\overline y}+
\gamma(\overline x),
\end{equation}

\noindent where $\overline x$ is the transformed of the point $x$.
Therefore, taking $\overline\gamma=\gamma$ and $\overline f(\overline y)=1$ we
can linearize the equation

\begin{equation}
\label{2b.5}
\ddot{y}+\alpha(x)f(y)\dot{y}+\beta(x)f(y)\int{f(y)\,dy}+ \gamma(x)f(y)=0,
\end{equation}

\noindent which transforms into

\begin{equation}
\label{2b.6}
{{\overline y}\,''}+\alpha(\overline x)\overline y{\,}'+
\beta(\overline x)\overline y +\gamma(\overline x)=0.
\end{equation}

An important physical problem of general relativity, concerning the motion of
expanding shear-free perfect fluids \cite{kus}, is governed by the ordinary
differential equation

\begin{equation}
\label{2b.7}
\ddot{y}=F(x)y^2,
\end{equation}

\noindent where $F(x)$ is an arbitrary function from which the equation of
state can be computed. A complete symmetry analysis of this differential
equation was given in \cite{wolf}. Here we see that it is contained in the set
of equations (\ref{2b.5}) when $\alpha(x)=0$, $\beta(x)=\frac{-3F(x)}{2}$,
$\gamma(x)=0$ and $f(y)=y^{1/2}$. Then, choosing $\overline f(\overline
y)=(\overline y)^{-1/2}$ in (\ref{7}-\ref{9.1}), the transformation of
variables is

\begin{equation}
\label{2b.8}
\overline y=\frac{y^3}{9},  \qquad {\overline x=\int{\frac{y^2}{3}\,dx}},
\end{equation}

\noindent and (\ref{2b.7}) becomes

\begin{equation}
\label{2b.9}
\overline y{\,}''=3F(\overline x),
\end{equation}

\noindent thus

\begin{equation}
\label{2b.10}
\overline y=\int{\left[\int{F(\overline x)\,d{\,}\overline x}\right]\,d{\,}
\overline x}+c_1\overline x+c_2,
\end{equation}

\noindent is the general solution of the simple linear equation
(\ref{2b.9}).

\section{Extended nonlocal transformation}

The integral in (\ref{16}) can be performed analytically and the general
solution $y=y(x)$ of (\ref{1}) obtained explicitly for a special set of
functions $f(y)$. For this purpose we generalize the nonlocal transformation
group defined by (\ref{7}-\ref{9.1}) extending it to

\begin{equation}
\label{17}
f_{11}(y)\,dy+f_{12}(y)\,dx=\overline f_{11}(\overline y)\,d{\,}\overline y+
\overline f_{12}(\overline y)\,d{\,}\overline x,
\end{equation}

\begin{equation}
\label{18}
f_{21}(y)\,dy+f_{22}(y)\,dx=\overline f_{21}(\overline y)\,d{\,}\overline y+
\overline f_{22}(\overline y)\,d{\,}\overline x.
\end{equation}

\noindent For simplicity we begin our investigations restricting ourselves to
the case $x=\overline x$, that is, $f_{21}=\overline f_{21}=0$,
$f_{22}=\overline f_{22}=1$ and requiring the invariant form (\ref{6}) to be
invariant under the remaining nonlocal transformation group, defined by
(\ref{17}-\ref{18}) with the above restrictions. Under these assumptions we
can write the nonlocal transformation as

\begin{equation}
\label{19}
\dot{\overline y}=p+q\dot{y},
\end{equation}

\noindent where the functions $p$ and $q$ are expressed in terms of the
functions $f_{11}, f_{12}, \overline f_{11}$ and $\overline f_{12}$. So, they
have a specific dependence on the variables $y$ and $\overline y$

\begin{equation}
\label{19.1}
p(y,\overline y)=\frac{f_{12}(y)}{\overline f_{11}(\overline y)}-
\frac{\overline f_{12}(\overline y)}{\overline f_{11}(\overline y)},
\end{equation}

\begin{equation}
\label{19.2}
q(y,\overline y)=\frac{f_{11}(y)}{\overline f_{11}(\overline y)}.
\end{equation}

\noindent Inserting (\ref{19}) in (\ref{6}) we get

$$
\frac{\ddot{y}}{f}+\alpha \dot{y}+\beta \int{f\,dy}+ \gamma =
\frac{q}{\overline f}\ddot {y}+
\left[\frac{\partial q}{\partial y}+
q\frac{\partial q}{\partial\overline y}\right]\frac{\dot y^2}{\overline f}+
$$
\begin{equation}
\label{22}
\left[\frac{\partial p}{\partial y}+q\frac{\partial p}{\partial\overline y}+
p\frac{\partial q}{\partial\overline y}\right]\frac{\dot{y}}{\overline f}+
\frac{p}{\overline f}\frac{\partial p}{\partial \overline y}+
\overline \alpha\left[p+q\dot{y}\right]+
\overline \beta \int{\overline f\,d{\,}\overline y}+ \overline\gamma,
\end{equation}

\noindent and comparing the coefficients of ${\dot{y}}^2$, we have

\begin{equation}
\label{20}
\frac{\partial q}{\partial y}+q\frac{\partial q}{\partial\overline y}=0,
\end{equation}

\noindent whose solution is

\begin{equation}
\label{21}
q(y,\overline y)=\frac{\overline y}{y}.
\end{equation}

\noindent Using (\ref{21}) and comparing the coefficients of $\ddot {y}$ we
easily find that $f=y$ and $\overline f=\overline y$. But, the comparisons of the
coefficients of $\dot {y}$ and the remaining terms give the equations

\begin{equation}
\label{23}
\alpha=\left[\frac{\partial p}{\partial y}+\frac{\overline y}{y}
\frac{\partial p}{\partial\overline y}+\frac{p}{y}\right]\frac{1}{\overline y}
+\overline \alpha \frac{\overline y}{y},
\end{equation}

\begin{equation}
\label{24}
\beta \int{y\,dy}+ \gamma =\frac{p}{\overline y}
\frac{\partial p}{\partial \overline y}+\overline \alpha p+
\overline \beta \int{\overline y\,d{\,}\overline y}+ \overline \gamma.
\end{equation}

\noindent The function $p$ that satisfies (\ref{23}) is given by

\begin{equation}
\label{25}
p(y,\overline y)=\frac{\alpha}{3}y\overline y -
\frac{\overline \alpha}{3}{\overline y}{\,}^2+h(y,\overline y),
\end{equation}

\noindent where the function $h(y,\overline y)$ satisfies the partial
differential equation

\begin{equation}
\label{25.1}
y\frac{\partial h}{\partial y}+
\overline y\frac{\partial h}{\partial\overline y}+h=0.
\end{equation}

\noindent It can be seen that the solutions of (\ref{25.1})
are given by $h=h_0/y$, where $h_0$ is an arbitrary function of the quotient
$\overline y/y$. So, the form of the solution for $p$ is

\begin{equation}
\label{25.2}
p(y,\overline y)= \frac{\alpha}{3}y\overline y -
\frac{\overline \alpha}{3}{\overline y}{\,}^2+\frac{h_0(\overline y/y)}{y}.
\end{equation}

\noindent Comparing (\ref{19.2}) with (\ref{21}) we have
$\overline f_{11}(\overline y)=1/\overline y$, and comparing (\ref{19.1})
with (\ref{25.2}), we obtain

\begin {equation}
\label{25.3}
h_0(\overline y/y)=c_1\frac{y}{\overline y}+c_2\frac{\overline y}{y}.
\end{equation}

\noindent Inserting (\ref{25.3}) in (\ref{24}) we find that $c_1=c_2=0$,
$\gamma+\beta c=\overline\beta{\,}\overline c+\overline \gamma$, and

\begin{equation}
\label{26}
\beta=\frac{2{\alpha}^2}{9},
\qquad{\overline \beta=\frac{2{\overline\alpha}{\,}^2}{9}}.
\end{equation}

\noindent Therefore, the final invariant form and the resulting nonlocal
transformation are

\begin{equation}
\label{27}
\frac{\ddot{y}}{y}+\alpha \dot{y}+\frac{{\alpha}^2}{9}y^2+\beta c+\gamma =
\frac{\ddot{\overline y}}{\overline y}+\overline \alpha \,\dot{\overline y}+
\frac{{\overline\alpha}{\,}^2}{9}{\overline y}{\,}^2+
\overline\beta{\,}\overline c+\overline\gamma,
\end{equation}

\begin{equation}
\label{28}
\frac{\dot{y}}{y}+\frac{\alpha}{3}y=\frac{\dot{\overline y}}{\overline y}+
\frac{\overline \alpha}{3}\overline y.
\end{equation}

\noindent In the particular case in which the invariant form (\ref{27})
vanishes, the l.h.s. gives rise to a nonlinear differential equation 

\begin{equation}
\label{28.1}
\ddot{y}+\alpha y\dot{y}+\frac{{\alpha}^2}{9}y^3+\gamma y=0,
\end{equation}

\noindent (where, without loss of generality we have taken $c=\overline c=0$,
so that, $\gamma =\overline \gamma$), that can be solved using the invariance
properties formulated above. To do this, we make $\overline\alpha =0$ on the
r.h.s. of (\ref{27}). Then, inserting its solution in (\ref{28}), it can be
integrated giving the general solution

\begin{equation}
\label{29}
y=\frac{3}{\alpha}\frac{2c_1 x+c_2}{c_1 x^2+c_2 x+c_3}, \qquad{\gamma =0}.
\end{equation}

\begin{equation}
\label{29.1}
y=\frac{3\sqrt\gamma}{\alpha}\frac{c_1\exp{\left(\sqrt\gamma x\right)}+
c_2\exp{\left(-\sqrt\gamma x\right)}}
{c_1\exp{\left(\sqrt\gamma x\right)}-c_2\exp{\left(-\sqrt\gamma x\right)}+c_3},
\qquad{\gamma\ne 0}.
\end{equation}

\noindent It can be seen that (\ref{28.1}) has eight Lie point symmetries and
it is equivalent to a second order linear differential equation under a point
transformation \cite{duarte}. On the other hand, for any other value of the
coefficient $\beta\ne\frac{2{\alpha}^2}{9}$, (\ref{28.1}) has two point Lie
symmetries and we cannot find a point transformation that cast it in a linear
equation \cite{duarte}. However, using the invariant form (\ref{10.1}) and the
transformation of variables (\ref{10}) for $f=y$, we have proved that
(\ref{28.1}) can always be linearized whatever the value of the coefficient of
$y^3$ is. Therefore, using the invariance properties of the form (\ref{6}) we
have obtained the same results that come by the Lie theory of symmetries. In
addition, we have linearized (\ref{28.1}) when it has less than eight Lie
point symmetries.

\section{Solution of new classes of differential equations}

Now, we are going to investigate the case when the invariant expression
(\ref{6}) vanishes, and we shall construct several important classes of
solvable second order nonlinear ordinary differential equations. To do this,
we must seek the nonlocal transformation defined by (\ref{19},\ref{21}) with
the condition that the invariant (\ref{22}) vanishes. This leads to the
equations that determine it

\begin{equation}
\label{32}
\alpha f=\frac{y}{\overline y}\left[\frac{\partial p}{\partial y}+
\frac{\overline y}{y}\frac{\partial p}{\partial\overline y}+
\frac{p}{y}\right]+\overline \alpha \overline f,
\end{equation}

\begin{equation}
\label{33}
\beta f\int{f\,dy}+ \gamma f=\frac{y}{\overline y}
\left[p\frac{\partial p}{\partial\overline y}+\overline\alpha p\overline f+
\overline\beta\,\overline f\int{\overline f\,d{\,}\overline y}+
\overline\gamma\overline f\right],
\end{equation}

\noindent and we shall show a set of functions $f,\overline f$ for which the
nonlocal transformation exists. The solution of (\ref{32}) can be obtained
writing

\begin{equation}
\label{33.1}
p(y,\overline y)=\alpha\overline yp_0(y)(y)+p_1(\overline y)+
p_2(y,\overline y),
\end{equation}

\noindent where each function satisfies

\begin{equation}
\label{35}
f=2p_0+yp_0',
\end{equation}

\begin{equation}
\label{33.2}
p_1'+\frac{p_1}{\overline y}+\overline \alpha \overline f=0,
\end{equation}

\begin{equation}
\label{33.3}
y\frac{\partial p_2}{\partial y}+
\overline y\frac{\partial p_2}{\partial\overline y}+p_1=0,
\end{equation}

\noindent where the $'$ indicates derivative with respect to the argument of
the function. Solving the system (\ref{35}-\ref{33.3}) and inserting their
solutions in (\ref{33.1}), we find the solution of (\ref{32}), that is:

\begin{equation}
\label{34}
p(y,\overline y)=\alpha\frac{\overline y}{y^2}\int{yf\,dy}
-\frac{\overline \alpha}{\overline y}\int{\overline f\overline y\,d{\,}\overline y}+
\frac{h_0(\overline y/y)}{y}.
\end{equation}

\noindent Comparing (\ref{34}) with (\ref{19.1}), the function
$h_0(\overline y/y)$ is given by (\ref{25.3}), but these terms can be
absorbed in a redefinition of the integration constants provided by the two
integrals of (\ref{34}). Then, without loss of generality we take them
equal to zero.

From (\ref{33},\ref{34}) we obtain the difficult integrodifferential
equation that satisfy the functions $f$ and $\overline f$. It reads

$$
-\frac{\alpha^2}{y^4}\left[\int{fy\,dy}\right]^2+\beta \frac{f}{y}\int{f\,dy}
+ \gamma\frac {f}{y}=
$$

\begin{equation}
\label{34.1}
-\frac{\overline\alpha{\,}^2}{\overline y{\,}^4}
\left[\int{\overline f\overline y\,d{\,}\overline y}\right]^2+
\overline\beta{\,}\frac {\overline f}{\overline y}
\int{\overline f\,d{\,}\overline y}+
\overline\gamma{\,}\frac {\overline f}{\overline y}.
\end{equation}

\noindent In what follows we shall show a set of functions $f,\overline f$ that are
solutions of this integrodifferential equation and construct three sets of
nonlinear differential equations that can be linearized and explicitly solved.

\subsection{Case a}

\noindent An interesting solvable equation set can be obtained when we choose
the functions $f,\overline f$ as:

\begin{equation}
\label{36}
f=by^n+k,  \qquad \overline f=\overline b{\,}{\overline y}{\,}^{\overline n}+
\overline k.
\end{equation}

\noindent Taking into account that the left hand side of (\ref{34.1})
depends of $y$ and its right hand side depends of $\overline y$, it must be a
constant. So, inserting the functions given by (\ref{36}) in
(\ref{34.1}) and after some algebra, it provides the constrains satisfied
by the parameters

\begin{equation}
\label{38}
\beta=\alpha ^2\frac{n+1}{(n+2)^2},  \qquad
\overline \beta=\overline\alpha{\,}^2\frac{\overline n+1}{(\overline n+2)^2},
\end{equation}

\begin{equation}
\label{42}
\beta k^2-\alpha^{2}\frac{k^2}{4}=\overline \beta\,{\overline k}{\,}^2-
{\overline\alpha}{\,}^2\frac{{\overline k}{\,}^2}{4}.
\end{equation}

\noindent In addition, the function $p(y,\overline y)$ is given by

\begin{equation}
\label{37.1}
p(y,\overline y)=\alpha\overline y\left[\frac{b}{n+2}y^n+
\frac{k}{2}\right]-\overline \alpha{\,}\overline y
\left[\frac{\overline b}{\overline n+2}\overline y^{\,\overline n}+
\frac{\overline k}{2}\right].
\end{equation}

\noindent Finally inserting (\ref{36}-\ref{42}) in the invariant form
(\ref{6}), we have

\begin{equation}
\label{40}
\ddot{y}+\alpha \left[by^n+k\right]\dot{y}+\beta \left[b^2\frac{y^{2n+1}}{n+1}
+bk\frac{n+2}{n+1}y^{n+1}+k^2y\right]=0,
\end{equation}

\begin{equation}
\label{41}
\ddot{\overline y}+\overline \alpha \left[{\,}\overline b{\,}\overline y^
{\,\overline n}+\overline k{\,}\right]\dot{\overline y}+
\overline \beta \left[\overline b^{\,2}\frac{\overline y^{\,2\overline n+1}}
{\overline n+1}+\overline b \,\overline k\frac{\overline n+2}{\overline n+1}y^
{\overline n+1}+\overline k^{\,2}\overline y\right]=0.
\end{equation}

\noindent Besides, from (\ref{19},\ref{21},\ref{37.1}) we obtain the
nonlocal transformation (\ref{17}) in invariant form

\begin{equation}
\label{43}
\frac{\dot y}{y}+\frac{\alpha by^n}{n+2}+\frac{\alpha k}{2}=
\frac{\dot{\overline y}}{\overline y}+
\frac{\overline\alpha{\,}\overline b{\,}{\overline y}^{\,\overline n}}
{\overline n+2}+\frac{\overline \alpha{\,}\overline k}{2},
\end{equation}

\noindent that links (\ref{40}) and (\ref{41}). To integrate these
equations we use their invariant property along with (\ref{38}-\ref{42}) and
analyze two different cases. In the first case, we choose $\overline b=0$,
$\overline \alpha =\alpha$, $\overline k=k$ and $\overline n=n$. Then,
$\overline \beta =\beta$ by (\ref{42}) and (\ref{41}) reduces to a
linear second order differential equation for $\overline y=\hat y$ with
constant coefficients

\begin{equation}
\label{44}
\ddot{\hat y}+\alpha k\dot{\hat y}+{\alpha}^2k^2\frac{n+1}{(n+2)^2}\hat y=0.
\end{equation}

\noindent Integrating (\ref{43}) for the above value of the parameter, we
obtain the general solution of (\ref{40})

\begin{equation}
\label{45}
y^n=\frac{n+2}{\alpha bn}\frac{\hat y^n}{\int{\hat y^n\,dx}},
\end{equation}

\noindent where $\hat y$ is any solution of (\ref{44}). In the second case,
when we choose $b=0$, $\alpha=\overline\alpha$, $k=\overline k$ and
$n=\overline n$, the (\ref{40}) reduces to (\ref{44}) for $y=\hat y$ and
the general solution of (\ref{41}) is

\begin{equation}
\label{47}
{\overline y}^{\,\overline n}=\frac{\overline n+2}
{\overline\alpha{\,} \overline b{\,}\overline n}
\frac{\overline{\hat y}{\,}^{\overline n}}
{\int{\overline{\hat y}{\,}^{\overline n}\,dx}},
\end{equation}

\noindent where $\overline {\hat y}$ is any other solution of (\ref{44}).
Inserting the general solution of the (\ref{40}) and (\ref{41}), given
by (\ref{45}) and (\ref{47}), in the nonlocal transformation
(\ref{43}), it can be integrated and the final relation between the
variables $y$ and $\overline y$, that transforms (\ref{40}-\ref{41})
one on each other, is

\begin{equation}
\label{48}
y\left[\int{\hat y^n}\,dx\right]^{\frac{1}{n}}\exp{\left(\frac{\alpha
k}{2}x\right)}
=\overline y\left[{\int{\overline{\hat y}{\,}^{\overline n}\,dx}}\right]^
{\frac{1}{\,\overline n}}
\exp{\left(\frac{\overline\alpha{\,}\overline k}{2}x\right)}.
\end{equation}

For the particular case $n=\overline n=-1$, we obtain $\gamma=\alpha^2b$
and $\overline\gamma=\overline\alpha^2\overline b$. All the remaining
equations (\ref{43})-(\ref{48}) can be applied for $n=-1$ and $\overline
n=-1$ because they do not depend explicitly of the parameters $\beta$,
$\overline\beta$, $\gamma$ and $\overline\gamma$.

In the next subsections we investigate other generalizations of
(\ref{40}-\ref{41}), that can be linearized and solved.

\subsection{Case b}

Writing the equations set (\ref{40}) and (\ref{41}) as

\begin{equation}
\label{b1}
F(\ddot{y},\dot{y},y)=0,        \qquad
\overline F(\ddot{\overline y},\dot{\overline y},\overline y)=0,
\end{equation}

\noindent a generalization of both equations can be done expressing them in
the following way,

\begin{equation}
\label{b2}
\frac {1}{y}F(\ddot{y},\dot{y},y)=\frac {1}{\overline y}
\overline F(\ddot{\overline y},\dot{\overline y},\overline y),
\end{equation}

\noindent which is invariant under the nonlocal transformation given by
(\ref{43}). It is easy to prove that the new functions

\begin{equation}
\label{b3}
\widetilde F(\ddot{y},\dot{y},y)=F(\ddot{y},\dot{y},y)+\delta y,   \qquad
\widetilde {\overline F}(\ddot{\overline y},\dot{\overline y},\overline y)=
\overline F(\ddot{\overline y},\dot{\overline y},\overline y)+
\delta\overline y,
\end{equation}

\noindent where $\delta$ is a constant parameter, also satisfy the invariant
condition (\ref{b2})

\begin{equation}
\label{b4}
\frac {1}{y}\widetilde F(\ddot{y},\dot{y},y)=\frac {1}{\overline y}
\widetilde{\overline F}(\ddot{\overline y},\dot{\overline y},\overline y).
\end{equation}

\noindent This {\it gauge symmetry} generates a new nonlinear equation that
can be linearized and solved. In fact, when the invariant in (\ref{b4})
vanishes, it gives rise to a set of equations that transform one on each other
under the same nonlocal transformation, these are:

\begin{equation}
\label{b5}
\ddot{y}+\alpha \left[by^n+k\right]\dot{y}+\beta \left[b^2\frac{y^{2n+1}}{n+1}
+bk\frac{n+2}{n+1}y^{n+1}+k^2y\right]+\delta y=0,
\end{equation}

\begin{equation}
\label{b6}
\ddot{\overline y}+\overline \alpha \left[{\,}\overline b\,\overline y^
{\,\overline n}+\overline k{\,}\right]\dot{\overline y}+
\overline \beta \left[\overline b^{\,2}\frac{\overline y^{\,2\overline n+1}}
{\overline n+1}+\overline b \,\overline k\frac{\overline n+2}{\overline n+1}y^
{\overline n+1}+\overline k^{\,2}\overline y\right]+\delta\overline y=0.
\end{equation}

\noindent In particular, to solve (\ref{b5}) we choose $\overline b=0$,
$\overline \alpha =\alpha$, $\overline k=k$ and $\overline n=n$ ($\overline
\beta =\beta$ by (\ref{42})) in (\ref{b6}). Then, it reduces to

\begin{equation}
\label{b7}
\ddot{\overline y}+\alpha k\dot{\overline y}+
\left[{\alpha}^2k^2\frac{n+1}{(n+2)^2}+\delta \right]\overline y=0.
\end{equation}

\noindent Inserting the solutions of (\ref{b7}) in (\ref{43})  and
integrating it for the selected parameters, we reduce (\ref{b5}) to
quadratures

\begin{equation}
\label{b8}
y=\left[\frac{n+2}{\alpha bn}\frac{\overline y\,^n}
{\int{\overline y\,^n\,dx}}\right]^{\frac{1}{n}}.
\end{equation}

\noindent For the particular case $\overline b=b=1$, $k=\overline k=0$,
$n=\overline n=1$ and $\delta =\gamma$, (\ref{b5}-\ref{b6}) reduce to
(\ref{28.1}), the variable transformation (\ref{43}) reduces to
(\ref{28}) and (\ref{48}) gives the relation between the variables $y$
and $\overline y$ that leaves invariant (\ref{27}).

\subsection{Case c}

There is an important result that can be deduced from (\ref{43}) when
$\overline\alpha=\alpha$ and $\overline k=k$, in this case the nonlocal
transformation (\ref{43}) is $k$-independent,

\begin{equation}
\label{50}
\frac{\dot y}{y}+\frac{\alpha by^n}{n+2}=
\frac{\dot{\overline y}}{\overline y}+
\frac{\alpha \overline b\,{\overline y}^{\,\overline n}}
{\overline n+2},
\end{equation}

\noindent and by (\ref{38}-\ref{42})

\begin{equation}
\label{51}
\overline n=n, \qquad  \overline n=\frac{-n}{n+1}.
\end{equation}

\noindent So, if we take $k(x)$ and $\delta (x)$ as functions of the
independent variable $x$ instead of constant parameters, then, there is no
change in the deduction of the variable transformation (\ref{50}), that
comes from (\ref{32}-\ref{33}). This means that the set of equations
(\ref{b5}-\ref{b6}) give rise to new solvable equations that transforms
between them by the nonlocal transformation (\ref{50})

\begin{equation}
\label{52}
\ddot{y}+\alpha \left[by^n+k(x)\right]\dot{y}+
\beta \left[b^2\frac{y^{2n+1}}{n+1}
+bk(x)\frac{n+2}{n+1}y^{n+1}+k^2(x)y\right]+\delta (x)y=0,
\end{equation}

\begin{equation}
\label{53}
\ddot{\overline y}+\alpha \left[{\,}\overline b\,\overline y^
{\,\overline n}+k(x)\right]\dot{\overline y}+
\beta \left[\overline b^{\,2}\frac{\overline y^{\,2\overline n+1}}
{\overline n+1}+\overline b \,k(x)\frac{\overline n+2}{\overline n+1}y^
{\overline n+1}+k^2(x)\overline y\right]+\delta (x)\overline y=0.
\end{equation}

\noindent For instance, to obtain the solutions of (\ref{52}) we
take $\overline b=0$ and $\overline n=n$ in (\ref{53}) and it becomes
a general homogeneous linear second order differential equation

\begin{equation}
\label{54}
\ddot{\overline y}+\alpha k(x)\dot{\overline y}+
\left[{\alpha}^2k^2(x)\frac{n+1}{(n+2)^2}+\delta (x)\right]\overline y=0,
\end{equation}

\noindent then, inserting the solutions of this equation in (\ref{b8}), we
reduce (\ref{52}) to quadratures.

\section{Conclusions}

We have introduced a new invariance concept that leads to classes of second
order nonlinear ordinary differential equations which are equivalent under
nonlocal transformations. These classes contain a second order linear ordinary
differential equation with constant coefficients. The parametric expression of
the solutions for an arbitrary function $f(y)$ and any values of the
parameters $\alpha$, $\beta$ and $\gamma$, has been found. Also, the case in
which these parameters are functions of the independent variable has been
investigated. Several important physical problems are mathematically described
by these equation classes. Many of these, arise in General Relativity when the
Einstein field equations are investigated for homogeneous, isotropic and
spatially flat cosmological models with no cosmological constant, or Bianchi
I-type metric with a variety of matter sources. Also, the probability
distribution function, which maximize the Fisher's information measure in the
generalized Statistical Mechanics, was found to satisfy (\ref{28.1}) for the
most interesting value $q=-1$ \cite{flavia}.

Taking $x=\overline x$ in the nonlocal transformation, and imposing the form
invariance of the general expression (\ref{6}), we have obtained a modified
Painlev\'e-Ince equation (\ref{28.1}). The nonlocal transformation of
variables and the general solution of these equations has been found. In this
case the equation has the eight dimensional group of Lie point group
symmetries SL(3,R) and this is the maximum number of point symmetries that a
second order differential equation can have. Other sets of new nonlinear
second order differential equations are generated, that can be linearized and
solved explicitly (\ref{40},\ref{b5},\ref{52}). It is also to be remarked
that, the use and application of the form invariance have lead to exact
solution of differential equations whose solution were unknown, in particular
for modified Painlev\'e-Ince equations and polinomical differential equations,
which usually appear in problem related with quantum effects in the very early
Universe, originated by the vacuum polarization terms and particle production
arising from a quantum description of matter, or when both of them are modeled
in terms of a classical bulk viscosity

In general, the problem of finding solutions of nonlinear ordinary
differential equations remains open. One direction along which one can proceed
is to reduce them to a linear ordinary differential equation. For instance,
when (\ref{1}) possesses eight-parameter Lie group it is linearizable by a
point transformation. On the other hand, the nonlocal transformation
(\ref{7}-\ref{9.1}) linearizes (\ref{1}) even when it has less symmetries.
Thus, it could mean that has more nonlocal symmetries. We conclude that it is
very interesting to study this kind of nonlocal transformations of variables
and their associated nonlocal symmetries, which have received up to  now
little attention. We shall continue exploring this subject in future papers.


\vskip 1cm

\noindent {\bf\large Acknowledgments}

\vskip .5cm

I want to acknowledge to F. Pennini and A.Plastino for sending me the result
of their preprint.

\newpage

\end{document}